\documentclass[11pt,a4paper]{article}
\usepackage{longtable}
\usepackage{amsmath}
\usepackage{graphicx}
\usepackage{bm}
\usepackage{footmisc}

\newcommand*{\no}{\noindent}
\newcommand*{\bea}{\begin{eqnarray}}
\newcommand*{\eea}{\end{eqnarray}}
\newcommand*{\be}{\begin{equation}}
\newcommand*{\ee}{\end{equation}}

\newcommand*{\pref}[1]{(\ref{#1})}

\newcommand*{\prefr}[2]{(\ref{#1}-\ref{#2})} 
\newcommand*{\nn}{\nonumber}

\newcommand*{\tr}{\mathrm{tr}}

\usepackage[square,comma,sort&compress,numbers]{natbib}

\title{On the Landau-gauge adjoint quark propagator}

\author{Daniel August, Axel Maas\\
Institute for Theoretical Physics, Friedrich-Schiller-University Jena,\\
Max-Wien-Platz 1, D-07743 Jena, Germany}

\begin{document}

\maketitle

\begin{abstract}
Quarks in the adjoint representation have been a subject of study for both conceptual and practical purposes. Conceptually, their differences when it comes to confining and chiral symmetry properties has long been suspected to hold important information on the relation of these two distinguished properties of QCD-like gauge theories. Practically, they have been studied as both a possibility to access finite density quark systems as well as candidate theories for technicolor in beyond-the-standard-model settings. The most elementary object describing such particles is their propagator, though it being gauge-dependent. Its properties in the minimal Landau gauge are investigated here both in the quenched and unquenched case for a range of lattice parameters using the Wilson formulation for the gauge group SU(2). It is found that the propagator shows pronounced differences to the case of fundamental quarks, especially towards the chiral limit.
\end{abstract}

\section{Introduction}

QCD belongs to a wider class of theories, in which gauge fields of a given Lie algebra interact with quark fields in some representation of the corresponding Lie group. Out of these possibilities, it is quarks in the adjoint representation which are quite interesting for several reasons.

In QCD it is found that in the quenched case the chiral and deconfinement transition coincides \cite{Karsch:2003jg}. Even in the unquenched case, where the transition becomes a cross-over, both transitions remain at the very least close by \cite{Cheng:2009zi,Cheng:2009be,Aoki:2009sc}. This has been found for all gauge groups investigated so far \cite{Lucini:2012gg,Holland:2003kg,Danzer:2008bk,Maas:2012wr}. This fact has been intriguing since a long time, and there were a multitude of investigations to understand whether there is a fundamental relation between both effects.

However, for adjoint quarks, i.\ e.\ quarks in the same representation as the gauge group, the situation is drastically different. There, both temperatures differ already in the quenched case by a gauge-group-dependent factor of four or more \cite{Bilgici:2009jy}. This remains true in the unquenched case  \cite{Karsch:1998qj,Engels:2005rr}. Again, this is true for all gauge groups investigated so far. The question is thus whether the relation of confinement and chiral symmetry breaking of the fundamental case is just purely coincidental, and not true for general representations, or whether adjoint quarks just work differently than fundamental quarks. That is particularly interesting as adjoint quarks do not have a sign problem at finite baryon density, and thus are an interesting proxy for QCD \cite{Kogut:2000ek,Hands:2000ei}. However, this makes only sense, if the mechanisms involved are not (too) different.

Once the adjoint quarks are dynamical, the theory shows again a very different behavior depending on the number of quark flavors, at least in the case of SU($N$) gauge groups. For a single Weyl fermion, the theory becomes supersymmetric. On the other hand, already at two flavors, the theory becomes likely quasi-conformal, i.\ e.\ walking, in the chiral limit \cite{Andersen:2011yj,Sannino:2009za}. It is therefore highly interesting for technicolor phenomenology \cite{Andersen:2011yj,Sannino:2009za,Hill:2002ap,Lane:2002wv}, and therefore has been investigated repeatedly \cite{Sannino:2009za,DelDebbio:2010hu,DelDebbio:2010hx,Bursa:2009we,DelDebbio:2009fd,Catterall:2008qk,Catterall:2007yx,Hietanen:2008mr,Hietanen:2009az,DeGrand:2009mt,DeGrand:2011qd,Lucini:2009an,Catterall:2009sb,Maas:2011jf,Patella:2012da,Andersen:2011yj,Hill:2002ap,Lane:2002wv}.

To understand these differences, a logical starting point is the most basic object describing these quarks: The quark propagator. For QCD, this propagator has been calculated many times with different methods, see e.\ g.\ \cite{Alkofer:2000wg,Fischer:2006ub,Bashir:2012cp,Kamleh:2007ud,Schrock:2011qp,Burgio:2012ph,Bonnet:2002ih,Zhang:2004gv,Parappilly:2005ei,Bowman:2005vx,Bowman:2002bm}. It shows clear signs of dynamical chiral symmetry breaking, and there are hints that it also shows that the quark is an unphysical particle due to the absence of a K\"allen-Lehmann representation, i.\ e.\ by positivity violation \cite{Alkofer:2003jj}. However, so far there are very few investigations for adjoint quarks \cite{Sannino:2009za,Aguilar:2010ad,Maskawa:1974vs,Fukuda:1976zb}.

Here, this situation will be improved on, using lattice simulations, for the case of the SU(2) gauge group. Since the quark propagator is gauge-dependent, this has to be done in a fixed gauge, which will be the minimal Landau gauge \cite{Maas:2011se}. After a brief discussion of the simulation details in section \ref{ssetup}, first the quenched and then the unquenched case will be investigated in sections \ref{squenched} and \ref{sunquenched}, respectively. The properties of the quark propagator will be detailed for a range of mass values and lattice parameters, in both momentum and position space. This results will be summarized together in the closing discussion in section \ref{ssummary}. These investigations continue and complement the analysis of the gluonic correlations functions of the unquenched theory in \cite{Maas:2011jf}

\section{Lattice setup}\label{ssetup}

\subsection{Configurations}

In the following the adjoint quark propagator will be investigated in the quenched and unquenched case. Since the unquenched configurations were obtained from the authors of  \cite{DelDebbio:2010hu,DelDebbio:2010hx,DelDebbio:2009fd,DelDebbio:2008zf}, the quenched simulations were chosen to match the unquenched ones, as has already been done in \cite{Maas:2011jf}. Since the unquenched ones are not made by us, we refer for the details of their creation to the original literature \cite{DelDebbio:2008zf}, and will detail here only their matching quenched case, as well as the additional calculations of the quark propagator. 

\begin{table}
\caption{\label{configs}The quenched configurations employed. $N_t$ is the temporal and $N_s$ the spatial extent of the lattice, and the total volume in lattice units is $N_t\times N_s^3$. The lattice spacing has been determined using the data of \cite{Fingberg:1992ju}, setting the string tension to $(440 $ MeV$)^2$. In all cases the quenched quark mass in physical units were $m=$0.01, 0.1, 0.5, 1, 2, and 10 GeV.}
\begin{tabular}{|c|c|c|c|}
\hline
$\beta$ & $a^{-1}$ [GeV] & Lattice sizes & Configurations \cr
\hline
2.221 & 0.985 & 8$^4$, 16$^4$, 24$^4$ & 1035, 667, 293 \cr
\hline
2.328 & 1.31 & 8$^4$, 16$^4$, 24$^4$ & 828, 850, 225 \cr
\hline
2.457 & 1.97 & 8$^4$, 16$^4$, 24$^4$ & 1035, 850, 199\cr
\hline
2.656 & 3.94 & 16$^4$, 24$^4$ & 750, 138\cr
\hline
\end{tabular}
\end{table}

Thus, in the quenched case the action was chosen to be the standard Wilson action, and configurations were created using the methods described in \cite{Cucchieri:2006tf}, i.\ e.\ with a combination of heat-bath and overrelaxation sweeps. The list of the lattice and configuration creation parameters for the quenched configurations are given in table \ref{configs}. For the unquenched case, the list of configurations can be found in \cite{Maas:2011jf}. Here, all but the configurations with a time extent of $N_t=64$ have been included, the latter being beyond the computational resources available to this project. For the quenched results, the usual QCD scale setting will be used \cite{Fingberg:1992ju,Cucchieri:2006tf}. However, for the dynamical case, this is a non-trivial problem \cite{DelDebbio:2010hu,DelDebbio:2010hx,DelDebbio:2009fd,DelDebbio:2008zf,Maas:2011jf}. Therefore results will be presented both in lattice units and in physical units chosen such that the lightest gau!
 ge-invariant $0^{++}$ state, which appears to be the glueball \cite{DelDebbio:2010hx}, will be given the mass of the, very likely, observed \cite{Aad:2012tfa,Chatrchyan:2012ufa} Higgs boson, i.\ e.\ 125 GeV, following the argumentation on the association of the Higgs mass with physical states of \cite{Frohlich:1981yi,Maas:2012tj}. This will be detailed more below in section \ref{sunquenched}.

Since the quark propagator is gauge-dependent, it is necessary to fix a gauge. To be compatible with previous investigations \cite{Maas:2011jf,Aguilar:2010ad}, and because of the general advantages \cite{Maas:2011se}, Landau gauge will be chosen. To cope with Gribov copies, this is further specified to the minimal Landau gauge \cite{Maas:2011se}. However, at least for scalar particles no significant influence on the propagator due to treatment of Gribov copies has been observed so far \cite{Maas:2010nc}, and thus this choice is expected to be of little relevance here. The actual gauge-fixing is then performed using the method described in \cite{Cucchieri:2006tf}.

\subsection{The adjoint quark propagator and lattice corrections}

\subsubsection{Continuum}

Due to the unbroken color symmetry the quark propagator in Landau gauge in the continuum in four Euclidean dimensions is parametrized by two scalar function $A$ and $B$ as
\bea
S^{ab}(p)&=&\delta^{ab}\frac{A(p^2)\gamma_\mu p_\mu+B(p^2)}{A(p^2)^2 p^2+B(p^2)^2}=\delta^{ab}Z(p^2)\frac{\gamma_\mu p_\mu+M(p^2)}{p^2+M(p)^2}\label{propagator}\\
Z(p^2)&=&\frac{1}{A(p)^2}\nn\\
M(p^2)&=&\frac{B(p^2)}{A(p^2)}\nn,
\eea
\no with the Euclidean Dirac matrices $\gamma_\mu$. The wave-function renormalization $Z$ and the mass function $M$ represent an alternative parametrization of the dressing functions $A$ and $B$. Especially, if a solution to the equation $M(-m^2)=-m^2$ exists the mass function determines the pole mass. However, even if a solution to the equation exists, it cannot be directly accessed with just the knowledge of the domain $p^2>0$, except if it would hold that $M(p^2)=M(-p^2)$, what is generically not the case.

Instead, to identify a possible mass, the Schwinger functions \cite{Alkofer:2003jj,Maas:2011se},
\bea
\Delta_v&=&\frac{1}{\pi}\int_0^\infty dp\cos(tp)\frac{Z(p^2)}{p^2+M(p^2)^2}\label{vschwinger}\\
\Delta_s&=&\frac{1}{\pi}\int_0^\infty dp\cos(tp)\frac{Z(p^2) M(p^2)}{p^2+M(p^2)^2}\label{sschwinger},
\eea
\no where $v$ refers to the vector part and $s$ to the scalar part, can be used. Just as ordinary space correlators \cite{Gattringer:2010zz}, these functions will be exponentially decaying at large times, with the decay constant being the pole mass, if such a pole mass exists \cite{Maas:2011se}. This will be investigated in sections \ref{sqps} and \ref{sups} for the quenched and unquenched case, respectively.

However, before this can be done, the propagator has to be renormalized, as it is logarithmically divergent in four dimensions. To renormalize the quark propagator requires in general two conditions, one for each tensor component, i.\ e.\ $A$ and $B$. These will be chosen to be $A_R(\mu)$=1 and $B_R(\mu)=m$, where $m$ is the tree-level mass. In the continuum, this can be achieved by two multiplicative renormalization factor $Z_A$ and $Z_B$, which obey in Landau gauge in this scheme $Z_A=Z_B$ \cite{Fischer:2006ub}, such that the $M$ function is renormalization-group-invariant. On the lattice, the situation will be different due to lattice artifacts.

\subsubsection{Lattice implementation}

Following \cite{DelDebbio:2008zf} for the lattice Dirac operator the Wilson operator with anti-periodic boundary conditions in time, but periodic boundary conditions in space, will be used,
\be
W(x,y)=-\frac{1}{2}\sum_\mu(1-\gamma_\mu)U_\mu^a\delta_{x+\mu y}+(m+4)\delta_{xy}\label{wo}.
\ee
\no $U_\mu^a$ are the links in the adjoint representation, which are obtained from the links $U_\mu$ in the fundamental representation by
\be
U_{\mu bc}^a=\frac{1}{2}\tr\left(\sigma^b U_\mu^+\sigma^c U_\mu\right)\no,
\ee
\no with the Pauli matrices $\sigma^a$. The corresponding lattice quark propagator in momentum space is then given by
\be
D_L=\delta^{ab}\frac{B_L(p^2)\left(m+\sum_\mu(1-\cos\left(\frac{2\pi P_\mu}{N_\mu}\right)\right)-A_L(p^2)i\sum_\mu\gamma_\mu\sin\left(\frac{2\pi P_\mu}{N_\mu}\right)}{B_L(p^2)^2\left(m+\sum_\mu(1-\cos\left(\frac{2\pi P_\mu}{N_\mu}\right)\right)^2+A_L(p^2)^2\left(\sum_\mu\sin\left(\frac{2\pi P_\mu}{N_\mu}\right)\right)^2}\nn,
\ee
\no where $P_\mu$ are the lattice momenta, which take the half-integer values $1/2,...,\linebreak(N+1)/2$ in time direction. This is the only direction which will be considered in the following, and therefore always all momenta will be tacitly taken to be in the time direction. The reasons for this is to make full use of the extended time direction of the unquenched configurations. In the quenched case, all lattices are symmetric, but the anti-periodic direction will still make a difference in a finite volume. To obtain $D_L$ requires to invert and Fourier-transform the Wilson operator \pref{wo}. This has been accomplished in a standard manner using a bi-conjugate gradient inversion on a plane-wave source for each momentum point, and subsequent projection onto the conjugate plane-wave. This code has been thoroughly checked for the free case, and for the case of pure global color symmetry, before going to the interacting case.

The ultimate aim is to determine an approximation to the continuum functions $A$ and $B$. Thus, the following functions will be determined, where the momentum is already chosen to be in the time direction,
\bea
pA_L&=&\frac{\tr \gamma_0 D_L}{(\tr\gamma_0 D_L)^2+(\tr D_L)^2}\nn\\
B_L&=&\frac{\tr D_L}{(\tr\gamma_0 D_L)^2+(\tr D_L)^2}\nn,
\eea
\no where $p$ is the relevant physical momentum component, $2\sin(\pi P/N_t)$. Of course, in the continuum limit this reduces to a trivial identity, $A_L=A$ and $B_L=B$. On a finite lattice, both are affected by lattice artifacts. Already for the free case, the exact results are \cite{Gattringer:2010zz}
\bea
A_L^\text{free}&=&\sin\left(\frac{2\pi P_0}{L}\right)\nn\\
B_L^\text{free}&=&m+1-\cos\left(\frac{2\pi P_0}{L}\right)\nn.
\eea
\no Thus, there is an additive, momentum-dependent lattice correction to the $B$ function and a multiplicative one to the $A$ function. The $A$ function can be corrected by division by an appropriate factor. The $B$ function can be corrected by either subtracting or normalizing the $B_L$ function. The latter option will be implemented here, as the $B$ function can then immediately be interpreted as the deviation from the free case. The corrected dressing functions are therefore
\bea
A&=&\frac{A_L}{A_L^\text{free}}\nn\\
B&=&\frac{B_L}{B_L^\text{free}}\label{bcorrection},
\eea
\no respectively.

The calculation of the Schwinger functions on the lattice afterwards is a straightforward discretization of the inverse Fourier transformations \prefr{vschwinger}{sschwinger}, taking into account the whole momentum range \cite{Maas:2011se}.

Statistical errors are calculated using bootstrap with 1000 resamplings and permitting asymmetric errors \cite{Cucchieri:2006tf}. For derived quantities, including the Schwinger functions, errors are propagated.

\section{Quenched}\label{squenched}

For the quenched results, to comply with the QCD orientation of this investigation, the scale is set by the string tension. It is given the value of $(440$ MeV$)^2$, using the results of \cite{Fingberg:1992ju} as input.

\subsection{Results in momentum space}

\begin{figure}
\includegraphics[width=\linewidth]{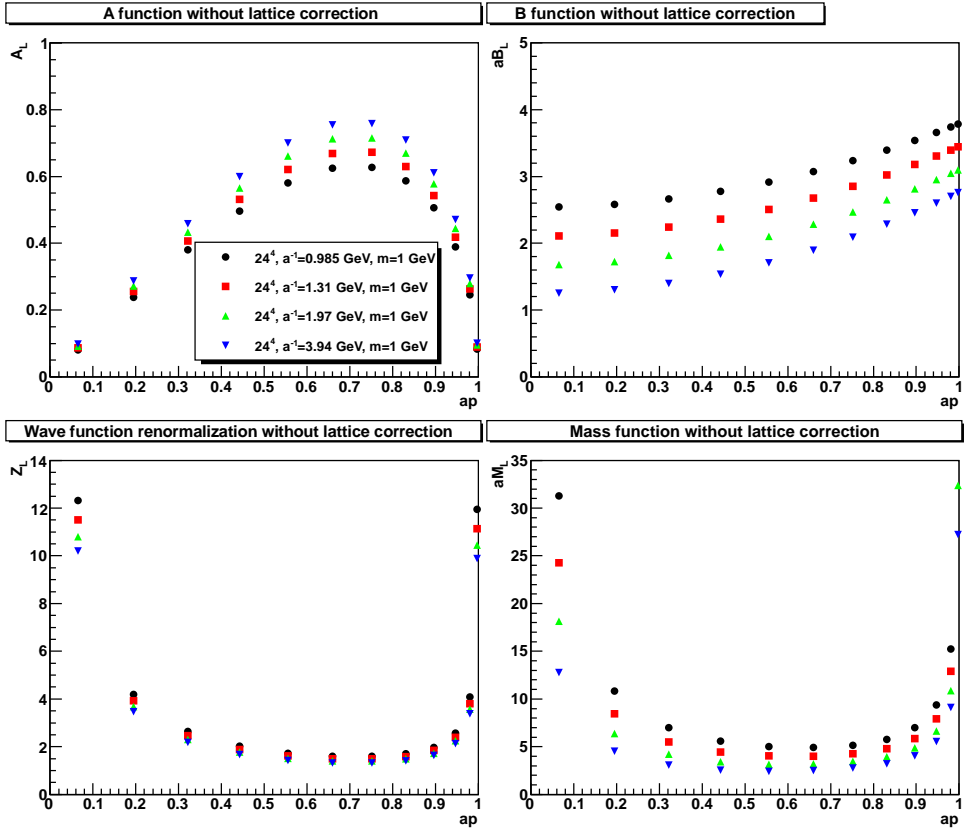}
\caption{\label{fig:wolat}The functions $A_L$ and $B_L$, as well as the derived quantities $Z_L=1/A_L$ and $M_L=B_L/A_L$ in lattice units for different lattice settings. Statistical errors here and afterwards are smaller than the box size, if not visible.}
\end{figure}

\begin{figure}
\includegraphics[width=\linewidth]{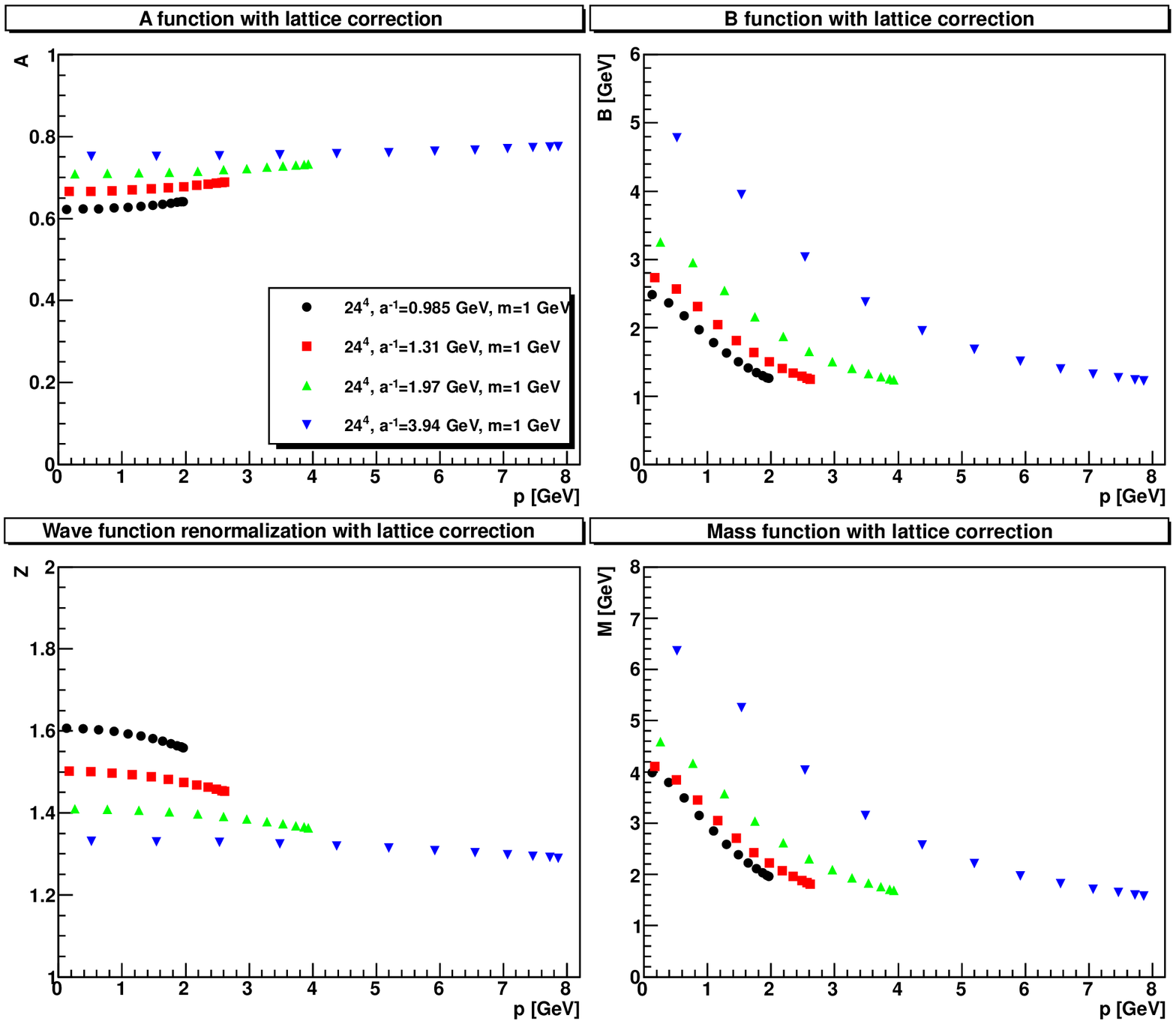}
\caption{\label{fig:wlat}The functions $A$ and $B$, corrected for leading lattice artifacts, as well as the derived quantities $Z=1/A$ and $M=B/A$.}
\end{figure}

To appreciate the importance of the lattice corrections, the uncorrected functions $A_L$ and $B_L$ are shown for several example lattice settings in figure \ref{fig:wolat}, as functions of the lattice momenta. Interestingly, the results do not deviate strongly from each other. In comparison, when including the lattice corrections, the situation changes, see figure \ref{fig:wlat}. It is immediately visible that the independence of the mass on the regulator is not obtained. This is likely due to the fact that Wilson fermions on a finite lattice introduce a mass shift, i.\ e.\ an additive mass renormalization, which is not present in the continuum \cite{Gattringer:2010zz}. Thus, the continuum renormalization prescription cannot be transferred. 

\begin{figure}
\includegraphics[width=\linewidth]{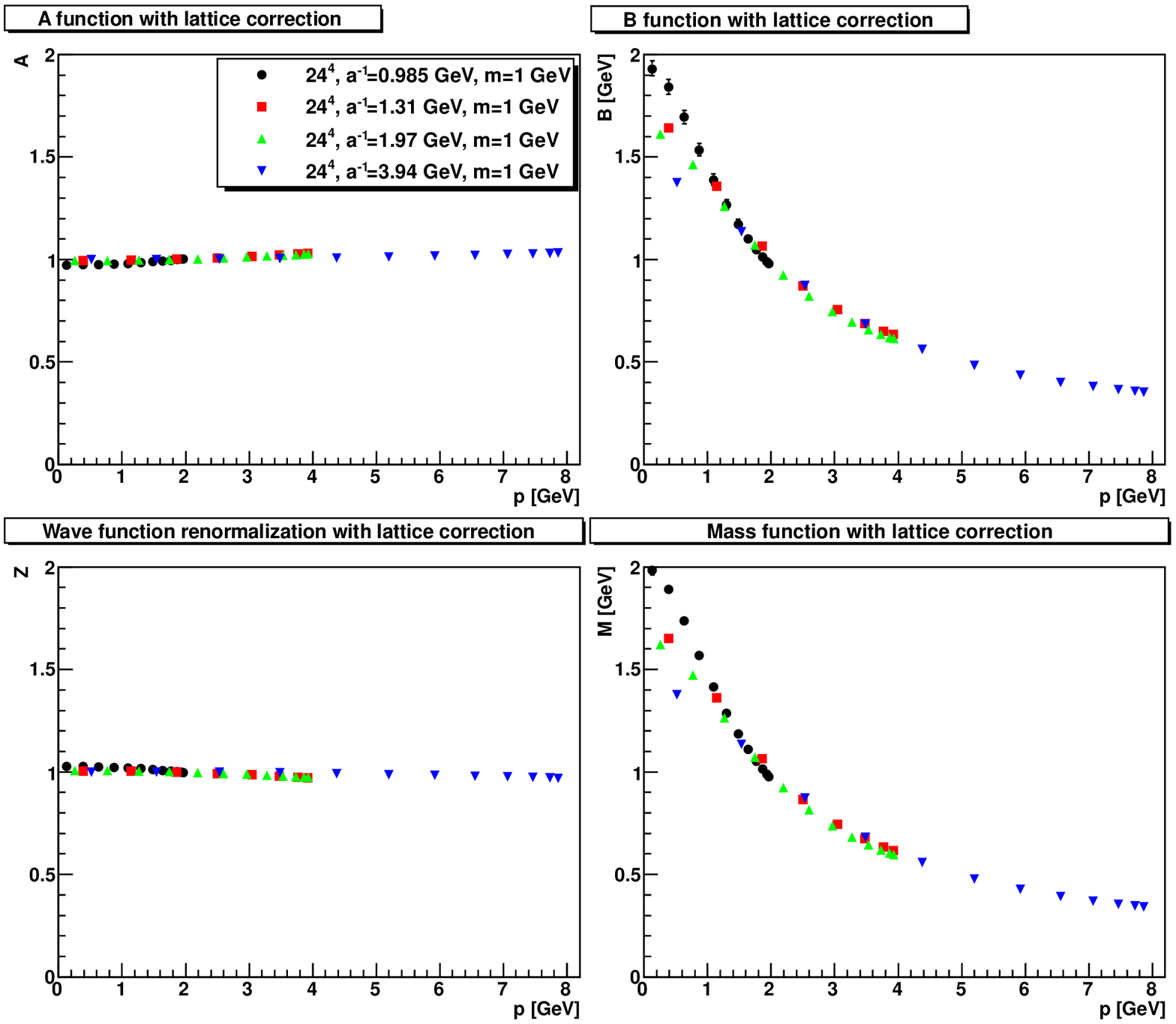}
\caption{\label{fig:ren}The renormalized functions $A$ and $B$, as well as the derived quantities $Z=1/A$ and $M=B/A$ is lattice units for different lattice settings.}
\end{figure}

As a consequence, the renormalization constants $Z_A$ and $Z_B$ are independent, and have to be included separately. They are determined using the renormalization conditions 
\bea
A(\mu)&=&1\nn\\
B(\mu)&=&m\label{brenormp},
\eea
\no with $\mu=1.9$ GeV. This value is chosen as it reduces discretization errors, since for all lattice spacings lattice momenta are sufficiently close to it available to permit a linear extrapolation of the dressing functions to $\mu$ to determine the multiplicative renormalization constants $Z_A$ and $Z_B$.

Doing so results in the transformation of figure \ref{fig:wlat} to \ref{fig:ren}, where the errors of the renormalization constants have been included by error propagation. The results show that there are still quite substantial lattice artifacts, which will be discussed in turn. For this purpose a bare quark mass is convenient, which is neither small nor large with respect to the lattice parameters. For this purpose, 1 GeV will be chosen.

\begin{figure}
\includegraphics[width=\linewidth]{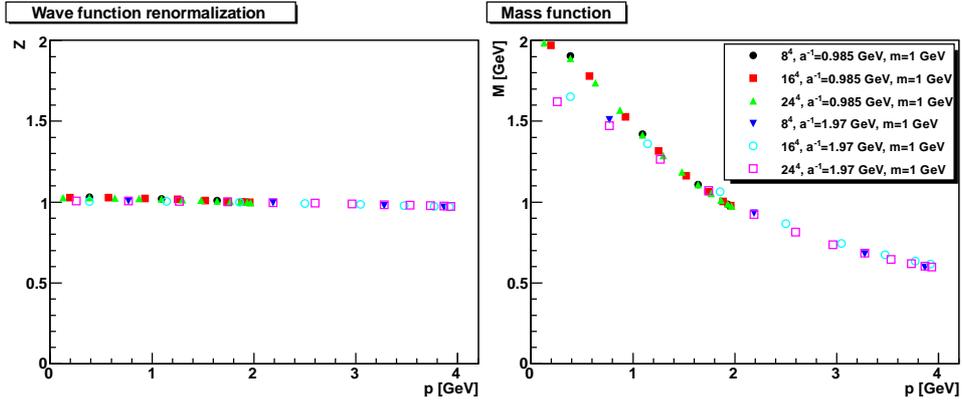}
\caption{\label{fig:v}The renormalized mass function and wave-function renormalization for two fixed lattice spacings and fixed mass and different physical volumes.}
\end{figure}

The first lattice artifact investigated is the dependence on the physical volume. This is shown in figure \ref{fig:v} for two different discretizations. Interestingly, there is almost no volume-dependence on the coarser lattice, while the results on a finer lattice show some effect, and in particular a large change compared to the coarser lattice. Comparing the case of a mass of 1 GeV to the cases of 0.5 and 2 GeV shows that this effect becomes smaller for a larger mass (the ratio at small momenta decreases from 1.2 to 1.1) while it increases for the smaller mass (to 1.4). This suggests that finite-volume effects become strong already at surprisingly large masses, given that $mL$ for 2 GeV and the largest volume is about 24. 

\begin{figure}
\includegraphics[width=\linewidth]{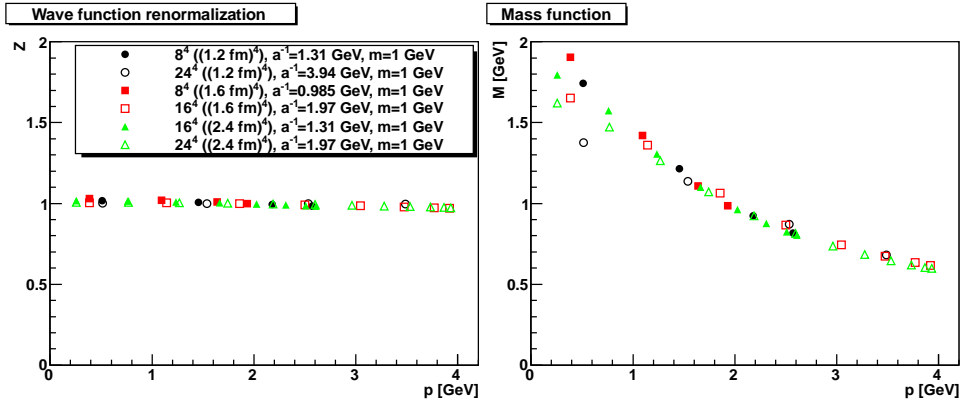}
\caption{\label{fig:a}The renormalized mass function and wave-function renormalization for three fixed physical volumes and fixed mass and different discretizations.}
\end{figure}

Also interesting is the comparison of discretization artifacts at fixed physical volume. These are shown in figure \ref{fig:a}. It is immediately visible that an improved discretization induces a stronger infrared suppression. However, this effect diminishes quickly with increasing physical volume. This effect becomes also smaller for larger mass and increases for smaller masses. This again emphasizes that for a mass of about 1 GeV, and even for 2 GeV, lattice artifacts play a substantial role.

The wave-function renormalization shows no pronounced dependency on either volume or discretization. However, a very close study shows that in a statistical significant way the wave-function renormalization becomes flatter with improved discretization, see figure \ref{fig:a}, while it is almost independent of volume. However, one should be wary that in the fundamental case different quark actions yield on similar lattice setups quite different wave function renormalizations \cite{Bowman:2002bm,Parappilly:2005ei,Zhang:2004gv}. It can therefore not be excluded that with a different discretization of the quark propagator the wave-function renormalization shows more substantial deviations from one. The impact of different quark propagator discretizations on the mass function is, however, by far not as severe.

Summarizing, the lattice artifacts for the lattice settings of table \ref{configs} for the employed masses are substantial, and induce especially at small momenta systematic errors exceeding 10\%. That is somewhat surprising, given the inherent mass-scale of the quarks, though gluonic correlators \cite{Maas:2011se} are similarly affected.

\begin{figure}
\includegraphics[width=\linewidth]{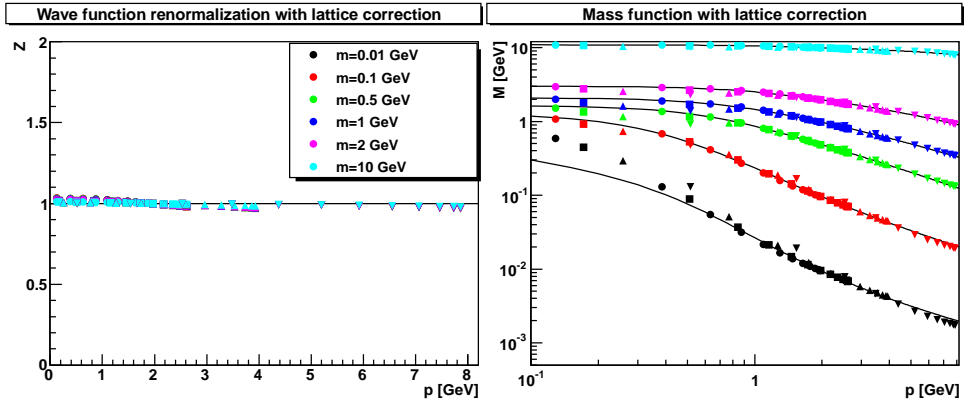}
\caption{\label{fig:fq}The wave function renormalization (left panel) and mass function (right panel) for the various masses and discretizations (circles are 0.2 fm, squares are 0.15 fm, triangles are 0.1 fm, and upside-down triangles are 0.05 fm), always from the 24$^4$ lattices. For the wave-function a constant one is also displayed, while for the mass-function fits of type \pref{qfit} are included by the full lines.}
\end{figure}

Keeping this in mind, the final results are presented in figure \ref{fig:fq}.

As already expected from the analysis of the lattice artifacts the wave-function renormalization does not deviate substantially from one. In fact, it is found that within errors all deviations from one depend on the lattice momentum rather than on the physical momenta, and diminish therefore at fixed physical momentum with better discretization. Thus, within errors, it appears that the wave-function renormalization is indeed constant. This is quite different from the case for fundamental quarks \cite{Fischer:2006ub,Kamleh:2007ud}, and somewhat surprising. However, it supports this assumption in several calculations using DSEs in the context of technicolor and conformal window studies \cite{Sannino:2009za,Maskawa:1974vs,Fukuda:1976zb}.

The mass-function shows a significant dependence on the physical momentum, and is only rather limited affected by the lattice artifacts. Only for the smallest mass of 0.01 GeV unsurprisingly significant lattice artifacts are seen, dominating especially at small momenta. It can thus be expected that only limited physical information can be obtained from this mass. However, even for the largest mass of 10 GeV, very few lattice artifacts are seen, which was not expected.

\begin{table}
\caption{\label{fitsq}Fit parameters for the quark mass function \pref{qfit}. Except for b and $\gamma$, which are dimensionless, all quantities are given in GeV.}
\begin{tabular}{|c|c|c|c|c|c|c|c|}
\hline
Mass & 2b & $\gamma$ & $\Lambda$ & $-\langle\bar{\Psi}\Psi\rangle^\frac{1}{3}$ & a & c & $M(0)$ \cr
\hline
0.01 & 0.2(4) & 1.9$^{-0}_{+1.3}$ & 1.2$^{+1.6}_{-0.6}$ & 0.12(7) & 0.003/0.01/2$\times 10^5$ & 1.6(1.2) & 0.18$_{-0.03}^{2.2}$ \cr
\hline
0.1 & 0.309(1) & 1.976(5) & 1.2$^{+0.1}_{-0.6}$ & 0.35(3) & 2.5$^{+0.3}_{-2.3}$ & 1.2$^{+0.2}_{-0.4}$ & 1.2(2) \cr
\hline
0.5 & 0.605(2) & 0.703(7) & 1.61(2) & 0.580(1) & 0.9(1) & 2.33(3) & 1.64(7) \cr
\hline
1 & 0.560(1) & 0.528(3) & 1.21(2) & 0.7239(8) & 0.90(4) & 2.01(3) & 2.08(8) \cr
\hline
2 & 0.6059(2) & 0.034(3) & 1.35(2) & 1.971(4) & 1.28(6) & 2.51(2) & 3.00(9) \cr
\hline
10 & 0.361(1) & 0.138(1) & 1.52(2) & 1.968(4) & 1.8(3) & 3.0(2) & 10.9(3) \cr
\hline
\end{tabular}
\end{table}

In all cases the mass function shows an infrared screening mass significantly larger than the mass implemented at the renormalization point. This is usually interpreted as a sign of contributions from chiral symmetry breaking \cite{Alkofer:2000wg}. To investigate this further, the mass function were fitted at large momenta by two possibilities \cite{Miransky:1986ib,Gusynin:1986fu}, the one the (up to explicit breaking) chiral symmetric case
\be
M(p)=M(\mu)\left(\omega\ln\frac{p^2}{\mu^2}+1\right)^{-\gamma},\label{pertform}
\ee
\no where $\omega$ and the anomalous mass dimension $\gamma$ were used as fit parameters. The alternative was the operator-product-expansion motivated form for the chirally broken case
\be
M(p)=\frac{2\pi^2\gamma}{3}\frac{-\langle\bar{\Psi}\Psi\rangle}{p^2\left(\frac{1}{2}\ln\frac{p^2}{\Lambda^2}\right)^{1-\gamma}},\nn
\ee
\no leaving the characteristic scale $\Lambda$, the chiral condensate $\langle\bar{\Psi}\Psi\rangle$, and $\gamma$ as free parameters. However, both forms do not satisfactorily fit the curves. This is achieved by using a more relaxed form
\be
M(p)=\frac{2\pi^2\gamma}{3}\frac{-\langle\bar{\Psi}\Psi\rangle}{p^{2b}\left(\frac{1}{2}\ln\frac{p^2}{\Lambda^2}\right)^{1-\gamma}},\nn
\ee
\no with anomalous exponent $b$. This form can be upgraded to a regularized version
\be
M(p)=\frac{2\pi^2\gamma}{3}\frac{-\langle\bar{\Psi}\Psi\rangle}{(p+a^2)^{2b}\left(\frac{1}{2}\ln\frac{p^2+c^2}{\Lambda^2}\right)^{1-\gamma}},\label{qfit}
\ee
\no with two further parameters $a$ and $c$, which then fits the quark mass function over the whole momentum range. The fit parameters, together with the screening mass $M(0)$ are given in table \ref{fitsq}. All fits were performed by dropping both the lowest and the two largest momentum points on all lattices, to remove the most dominant lattice artifacts in the fits. Of course, $\gamma$ and $\langle\bar{\Psi}\Psi\rangle$ in the fit form \pref{qfit} can no longer be strictly interpreted as the anomalous dimension and the chiral condensate, except when $b\approx 0$, which is not the case. 

As is visible, the strong finite-volume effects on the smallest mass makes the fit completely unreliable, while for the other masses quite stable results can be obtained. In particular, the physically most interesting parameters are most robust, while the regularization parameters $a$ and $c$ show somewhat stronger fluctuations, especially for smaller masses. It is interesting that the would-be anomalous mass dimension shows a pronounced dependence on the mass, moving towards the unitarity limit of -2 for decreasing mass.

The results show that the effective screening mass $M(0)$ is indeed about 1 GeV larger than the bare mass. Interpreting this as a chiral scale, this scale is thus much larger than the one for fundamental quarks, which is of order 300 MeV \cite{Fischer:2006ub}. If one takes the value of the chiral condensate seriously, which is probably not adequate since the exponent $b$ is far from 1, this would suggest rather a chiral scale similar to the one of fundamental quarks. To better understand this, it is worthwhile to investigate the quark propagator in configuration space, i.\ e.\ its Schwinger functions \prefr{vschwinger}{sschwinger}.

\subsection{Results in position space}\label{sqps}

\begin{figure}
\includegraphics[width=\linewidth]{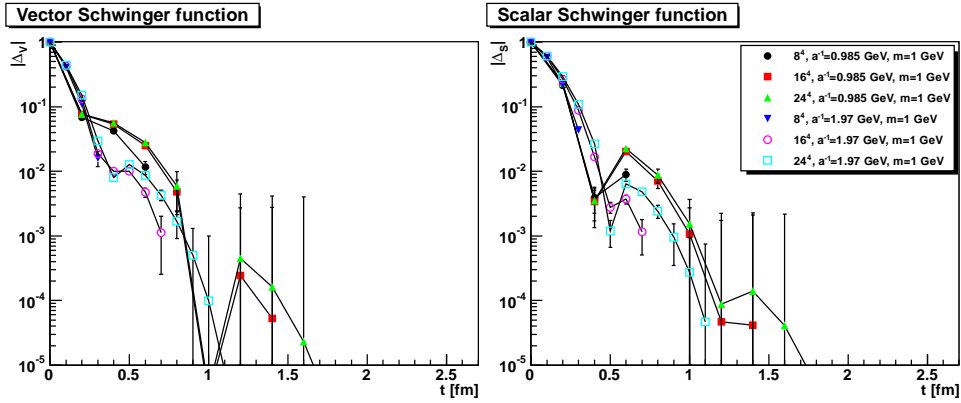}
\caption{\label{fig:vp}The vector (left panel) and scalar (right panel) Schwinger functions for two fixed lattice spacings and fixed mass and different physical volumes. Lines are to guide the eye.}
\end{figure}

\begin{figure}
\includegraphics[width=\linewidth]{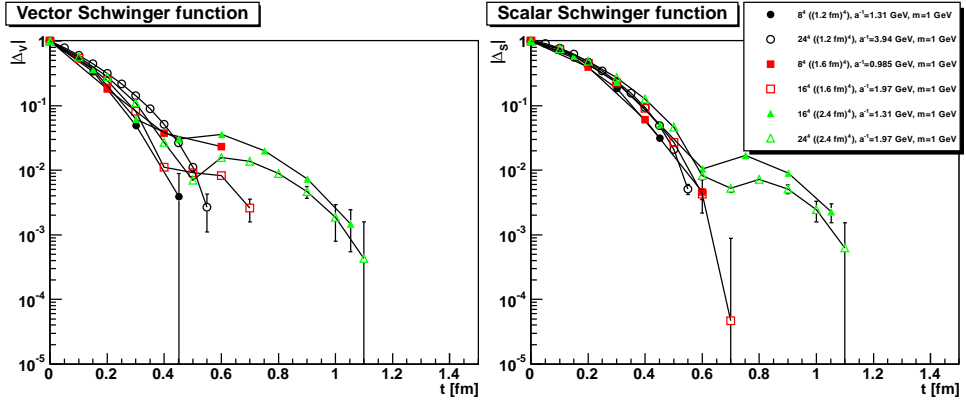}
\caption{\label{fig:ap}The vector (left panel) and scalar (right panel) Schwinger functions for three fixed physical volumes and fixed mass and different discretizations. Lines are to guide the eye.}
\end{figure}

Though the momentum-space results do not show substantial lattice artifacts once the most extreme momenta are neglected, this can change in position space. Therefore, figures \ref{fig:vp} and \ref{fig:ap} show the analogues of figures \ref{fig:v} and \ref{fig:a} in position space, i.\ e.\ the dependence on volume and discretization, respectively. The severeness of volume artifacts is seen to be stronger the better the lattice discretization. While for the coarse lattice this is not too severe, it is very substantial for the finer lattice. The discretization effects seen in figure \ref{fig:ap} are also sizable. However, their size strongly depend on the volume, and they become smaller the larger the volume. Thus, the most important requirement is a sufficiently large volume. However, for all lattice parameters available, sizable effects remain, which will distort any quantitative results at least by 10\%.

\begin{figure}
\includegraphics[width=\linewidth]{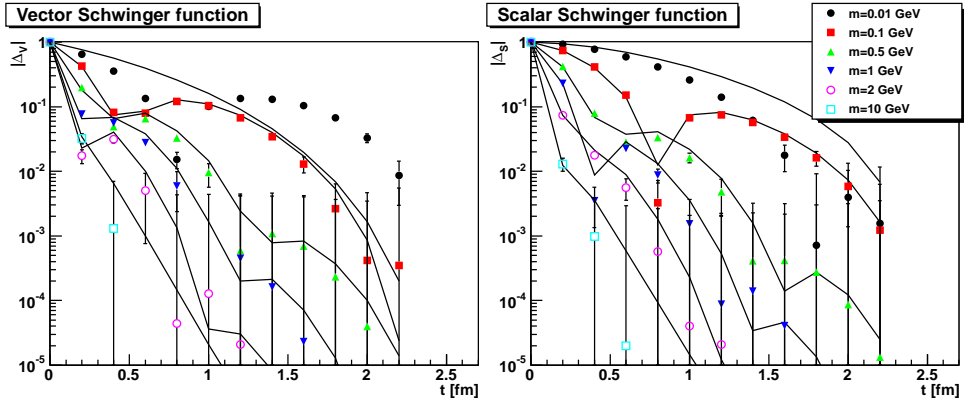}
\caption{\label{fig:schwinger}The vector (left panel) and scalar (right panel) Schwinger functions for the different masses from the largest volume. Lines are the lattice Schwinger functions of the corresponding fits \pref{qfit}.}
\end{figure}

Keeping these systematic effects in mind, the best choice appears to use the largest physical volume. The behavior for the different masses is then shown in figure \ref{fig:schwinger}. The first observation is the presence of a dip in all cases, except for the largest mass. This dip in the plotted absolute value of the Schwinger functions originate from a sign change in the Schwinger function itself, thus implying a positivity violating spectral function \cite{Alkofer:2003jj,Maas:2011se}. Hence, quenched adjoint quarks are not part of the physical spectrum, similar to gluons \cite{Maas:2011se}. The position of this dip depends strongly on the mass, and it is generally at shorter distances for the vector Schwinger function than for the scalar Schwinger function.

The corresponding (lattice) Schwinger functions of the fits \pref{qfit} describe the behavior rather well, except for the lightest mass. Interestingly, they signal the presence of more zero crossings, though this cannot be unambiguously confirmed by the lattice data itself due to the large statistical fluctuations. Such an oscillatory behavior would indicate conjugate complex poles as the analytical structure \cite{Alkofer:2003jj,Maas:2011se}. This would be in-line with the commonly assumed structure for the fundamental quark propagator \cite{Alkofer:2000wg,Fischer:2006ub}.

\section{Unquenched}\label{sunquenched}

Before discussing the results for the unquenched adjoint quark propagator some particularities of the unquenched configurations have to be addressed.

The first one is with regard to the scale. There is no (known) equivalent to the unquenched theory in nature. The results on the bound state spectrum \cite{DelDebbio:2010hu,DelDebbio:2010hx,Bursa:2009we,DelDebbio:2009fd} furthermore suggest that the physics is completely different from the QCD case, with the lightest glueball always being lighter than the lightest hadron. Especially, breaking of chiral symmetry in the chiral limit could not be established, and the theory is expected to be walking or, possibly, quasi-conformal or even conformal in the chiral limit \cite{Sannino:2009za,Andersen:2011yj,DelDebbio:2009fd}.

This makes the scale setting quite arbitrary. Based on technicolor phenomenology, in the previous work \cite{Maas:2011jf} the lightest state in the channel $0^{++}$, the dilaton, being here the scalar glueball \cite{DelDebbio:2010hx}, has been used in to set the scale, assuming a mass of 2 TeV. This scale provided a rather smooth behavior of gluonic correlators with the quark mass, and appears therefore reasonable. It also coincides, within errors, with the scale determined using the conventional fundamental string tension. It is this approach which will be followed her to set a physical scale. In addition, to avoid any bias introduced by this scale setting, all final results will also be shown in lattice units.

However, the value of 2 TeV appears questionable. Based on the considerations in \cite{Frohlich:1981yi,Maas:2012tj}, it should be expected that the scalar glueball/dilaton state actually mixes with the observed \cite{Aad:2012tfa,Chatrchyan:2012ufa} Higgs-like particle, if it has not even to be identified with it. Based on these considerations, this state should therefore be attributed rather a mass of 125 GeV, which will be done here.

Finally, there is the question of the renormalization scheme. Without explicit referral to the quarks, previously it sufficed to just use the bare quark mass \cite{Maas:2011jf}. However, for the unquenched case most of the bare masses are actually negative \cite{DelDebbio:2010hu,DelDebbio:2010hx,Bursa:2009we,DelDebbio:2009fd}, especially towards the chiral limit, and therefore unsuitable for the renormalization prescription. To address this problem, the PCAC quark mass \cite{DelDebbio:2010hu}, being a positive quantity, will be used here instead of the bare mass for the renormalization condition \pref{brenormp}. Since this mass is only available over a limited range, it was extrapolated/interpolated linearly for all other cases, particularly towards large bare quark masses. This mass was also used for the lattice correction \pref{bcorrection}. Since in the limit $a\to 0$ this choice drops out in the renormalization procedure, any deviation from an optimal choice to remove la!
 ttice artifacts only surfaces again as a lattice correction.

Note that the configurations available to us are only at one fixed bare coupling $\beta=2.25$, though with varying bare quark mass, and only for a very limited range of volumes. This limits the amount of systematic analysis possible severely compared to the quenched case.

\subsection{Impact of the wrong phase}

A serious problem surfacing is that a spatial center-breaking transition is observed \cite{DelDebbio:2010hx} when decreasing the bare lattice mass towards the chiral limit of $am\approx -1.2$ \cite{DelDebbio:2010hu}. It is observed for a spatial extent of $N_s=8$ at bare quark masses below $am=-0.975$, for $N_s=12$ below $am=-1.05$, and for $N_s=16$ to set in between $am=-1.05$ and $am=-1.15$ \cite{DelDebbio:2010hu}. The spatial dependence suggests that changing the bare quark mass at fixed gauge coupling $\beta$ also changes the lattice spacing, and especially reducing it toward the chiral limit. This is supported by other observations in the gluonic sector \cite{Maas:2011jf}. Thus the physical volume shrinks when lowering the bare quark mass at fixed $\beta$, and at a certain physical extent a phase transition occurs, just like in Yang-Mills theory. Therefore, in the following this transition will be assumed to be a finite-volume artifact.

\begin{figure}
\includegraphics[width=\linewidth]{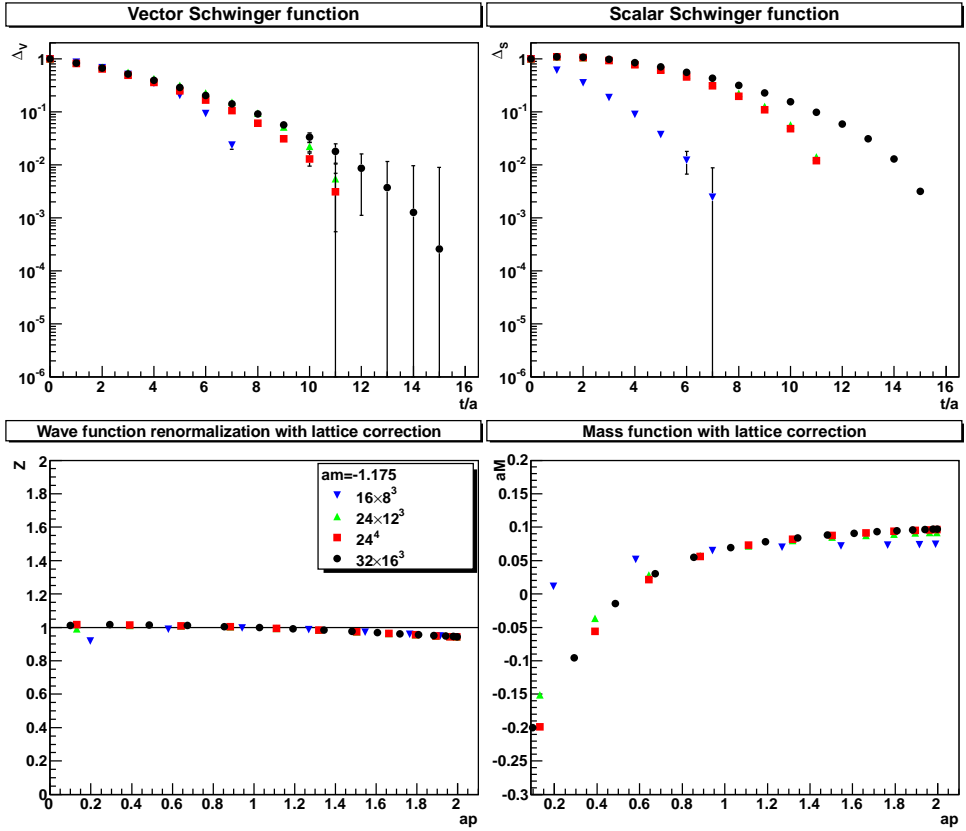}
\caption{\label{fig:artifact}The vector (top left panel) and scalar (top right panel) Schwinger functions and the wave function renormalization (bottom left panel) and mass function (bottom right panel) for different volumes for $am=-1.175$, i.\ e.\ a PCAC mass of 0.0678.}
\end{figure}

This did neither influence significantly gluonic correlators \cite{Maas:2011jf}, nor did it distort the physical spectrum in lattice units \cite{DelDebbio:2010hu,DelDebbio:2010hx}. In contrast, and thus surprisingly, we find here that it does alter the quark dressing functions, especially the $B$ function, qualitatively. Especially, it induces a sign change at a finite momentum. This is illustrated in figure \ref{fig:artifact}. While the vector Schwinger function and the wave-function renormalization appear still mostly normal, this is dramatically different for the scalar part. The scalar Schwinger function increases at first. This is not yet a problem, as it does so also in the quenched case after becoming negative, see figure \ref{fig:schwinger}. The consequence for the mass function is much more severe: It has a sign change. This originates from a sign change of the uncorrected $B$ function. This behavior first increases with volume from $N_s=8$ (where the function is st!
 ill positive) to $N_s=12$. Reducing the aspect ratio from $1:2$ to $1:1$, this effect is counteracted. It is also reduced when going to $N_s=16$. The latter is expected for a finite-volume artifact. Unfortunately, there a not enough different volumes available to check this explicitly close to the phase transition. Therefore, it can only remain a conjecture that this is a pure finite-volume artifact.

When instead of the bare quark mass a sufficiently larger quark mass is used when determining the quark propagator, i.\ e.\ a larger bare valence quark mass than sea bare quark mass, the $B$ function becomes once more positive. However, since it is not clear what kind of physics such a partially quenched theory actually describes, this will not be explored further here.

This serious lattice artifact therefore prevents any physical interpretation of the quark propagator inside the wrong phase, and limits the following investigation to the cases with volumes large enough to have at least the possibility to be in the infinite-volume phase.

\subsection{Results in momentum space}

\begin{figure}
\includegraphics[width=\linewidth]{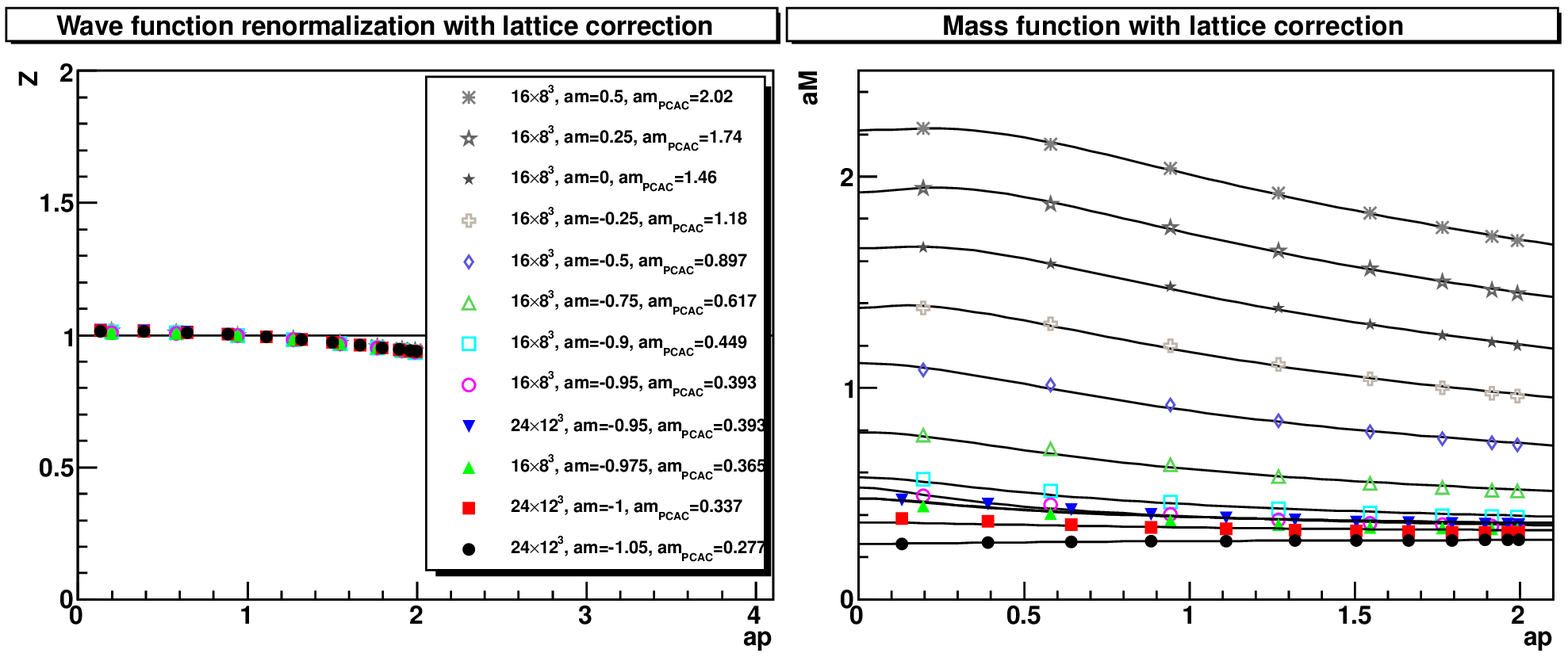}
\includegraphics[width=\linewidth]{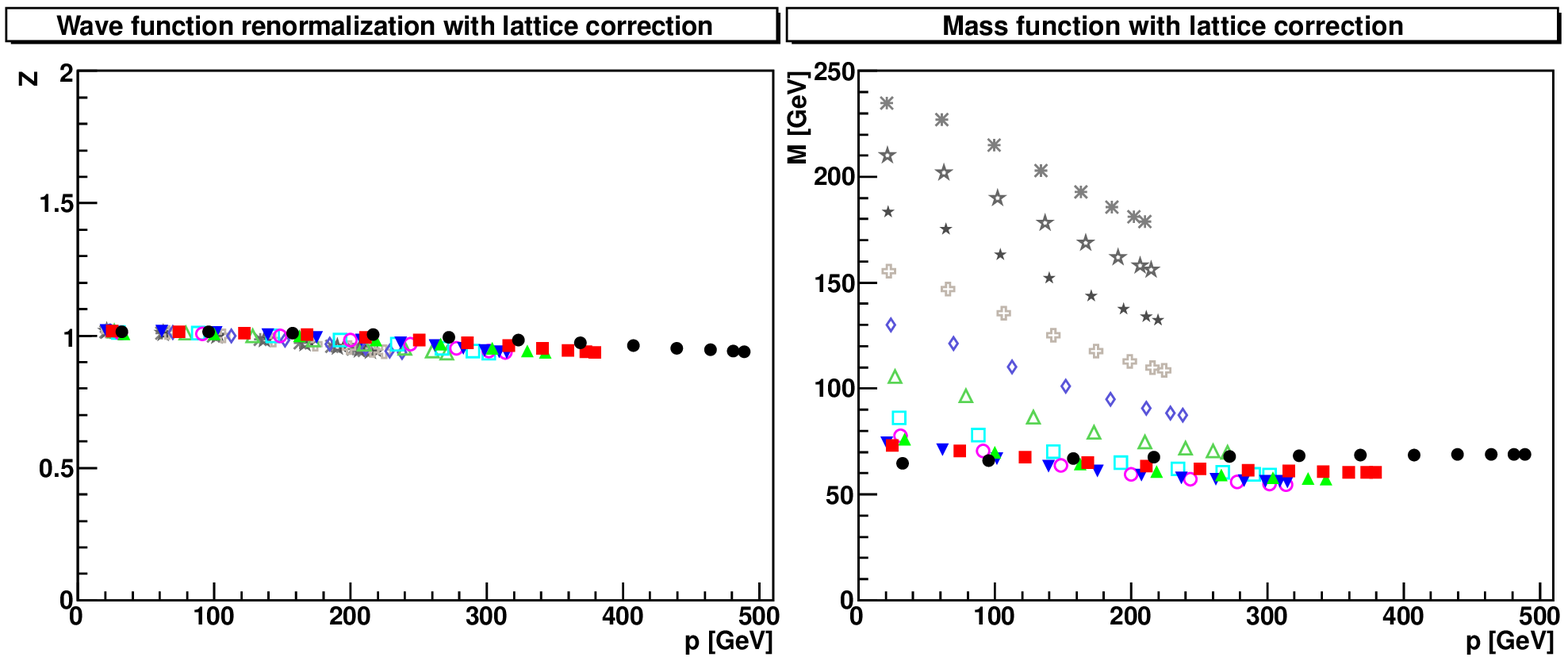}
\caption{\label{fig:tcm}The wave function renormalization (left panels) and mass function (right panels) in lattice units (top panels) and physical units (bottom panels) for different masses and volumes. For the case of lattice units also fits of type \pref{qfit} are shown. Renormalization is performed at $a\mu=1$.}
\end{figure}

\begin{table}
\caption{\label{tab:tcfits}Fit parameters for the unquenched quark mass function \pref{qfit}. Dimensionful units are given in GeV.}
\begin{tabular}{|c|c|c|c|c|c|c|c|c|c|}
\hline
$am_0$ & $N_s$ & $am_\text{PCAC}$ & $m_\text{PCAC}$ & 2b & $\gamma$ & $aM(0)$ & $M(0)$ & $-a\langle\bar{\Psi}\Psi\rangle^\frac{1}{3}$ & $-\langle\bar{\Psi}\Psi\rangle^\frac{1}{3}$ \cr
\hline
0.5 & 16 & 2.02 & 214 & 0.18(2) & 0.967(6) & 2.22(5) & 235(6) & 0.691(2) & 73.2(3) \cr
\hline
0.25 & 16 & 1.74 & 188 & 0.18(2) & 0.972(8) & 1.9(2) & 205(22) & 0.656(3) & 70.9(4) \cr
\hline
0 & 16 & 1.46 & 161 & 0.18(2) & 0.980(8) & 1.7(1) & 187(11) & 0.618(2) & 68.0(3) \cr
\hline
-0.25 & 16 & 1.18 & 133 & 0.17(2) & 0.991(7) & 1.41(2) & 159(3) & 0.573(1) & 64.7(2) \cr
\hline
-0.5 & 16 & 0.897 & 108 & 0.15(2) & 0.99(2) & 1.09(3) & 131(4) & 0.517(7) & 62.0(9) \cr
\hline
-0.75 & 16 & 0.617 & 83.9 & 0.11(2) & 0.97(2) & 0.78(7) & 106(10) & 0.446(5) & 60.7(7) \cr
\hline
-0.9 & 16 & 0.449 & 68.2 & 0.06(3) & 0.96(2) & 0.50(7) & 76(11) & 0.41(1) & 62(2) \cr
\hline
-0.95 & 16 & 0.393 & 62.1 & 0.01(7) & 0.93(6) & 0.51(17) & 81(28) & 0.37(2) & 58(4) \cr 
\hline
-0.95 & 24 & 0.393 & 62.1 & 0.02(3) & 0.96(2) & 0.46(6) & 73(10) & 0.38(1) & 60(2) \cr
\hline
-0.975 & 16 & 0.365 & 62.8 & 0.04(11) & 0.89(11) & 0.34(18) & 58(31) & 0.36(2) & 62(4) \cr
\hline
-1 & 24 & 0.337 & 64.0 & 0.00(2) & 0.97(2) & 0.35(6) & 67(11) & 0.366(6) & 70(2) \cr
\hline
-1.05 & 24 & 0.277 & 67.9 & 0.02(4) & 1.00(3) & 0.26(5) & 64(13) & 0.346(6) & 85(15) \cr
\hline
\end{tabular}
\end{table}

The results for the wave function renormalization and mass function in both lattice and physical units in the relevant bare mass range are shown in figure \ref{fig:tcm}. Again, a fit with the form \pref{qfit} is possible, and the parameters are shown in table \ref{tab:tcfits}. However, since the exponent $b$ is close to zero for small quark masses, and $\gamma$ is always close to one, results for the constants $a$, $c$, and $\Lambda$ are quite unreliable and very sensitive to statistical fluctuations. They are therefore not listed, since their usefulness is very limited.

The wave-function renormalization is, as in the quenched case, for all practical purposes indistinguishable from one. For the mass function, however, there are quite a number of differences compared to the quenched case. 

The first is that the mass function is much slower decreasing towards the ultraviolet. Especially, for smaller quark masses the exponent $b$ is essentially zero, while the anomalous exponent $\gamma$ is close to one. In fact, for the smallest quark masses accessible without phase transition the mass function is essentially constant. Also, the effective mass $M(0)$ is much less enhanced compared to the mass function at the renormalization point. This ratio is, within errors, even one for the smallest quark masses. This indicates that for small quark masses there is no spontaneous chiral symmetry breaking left, and only explicit breaking remains. These statement holds true independent of the units.

In physical units, in contrast to the case of the gluonic correlation functions \cite{Maas:2011jf}, the mass function does not collapse onto a universal curve. Given that the amount of chiral symmetry breaking is reduced with decreasing mass, and therefore also the effective mass $M(0)$ decreases quicker, this is not surprising. Interestingly, the chiral condensate in physical units stays more or less constant, as do the PCAC masses in physical units, as soon as the PCAC mass reaches a certain level. At this level, the curves again collapse to a certain degree to the same curve.

These results are not easily interpreted. It appears that moving to smaller masses reduces the amount of spontaneous chiral symmetry breaking, until it essentially vanishes. This would be in line with the expectation that the theory develops a (quasi-)conformal infrared fixed point in the chiral limit  \cite{DelDebbio:2010jy,Deldebbio:2010ze}.

However, whether this is indeed an effect of reaching the chiral limit, or whether this is a finite-volume artifact due to the vicinity of the spatial phase transition is less easy to understand. Since the effect seems to be rather volume insensitive at fixed quark mass, this may be possibly assigned to the former option. However, without much more systematic investigations, this could also be wishful thinking. Thus, this result is more an encouragement to extend the systematics, rather than a final result.

With the exponent $b$ being close to zero, the form of the propagator is close to the form \pref{pertform}. This would permit to interpret $\gamma$ at small quark masses indeed as the anomalous dimension, which would then be unexpectedly large, and close to one. This is in stark contrast to indirect determinations possible \cite{DelDebbio:2010jy,Deldebbio:2010ze} from gauge-invariant observables, like the hadronic spectrum \cite{DelDebbio:2010hu} or the spectrum of the Dirac operator \cite{Patella:2012da,Deldebbio:2010ze}. Whether this is a lattice artifact, a (residual) scheme dependency given that $Z_A\neq Z_B$ due to lattice artifacts, or has another reason is a question which has to be explored.

However, there is, of course, a certain arbitrariness in the fit form \pref{qfit}. Given that the largest accessible momenta are not too large compared to all other scales, the regularization can influence the result. If, e.\ g., a fit form generalizing \pref{pertform} of type
\be
M(p)=a\left(\omega\ln\frac{p^2}{\Lambda^2}+k\right)^{-\gamma_p}\nn
\ee
\no is used instead, the obtained anomalous exponent $\gamma_p$ is found to be of order 0.2-0.3 for small quark masses \cite{August:2012ma}. This is in much better agreement to other available determinations, e.\ g.\ \cite{DelDebbio:2010hu,Patella:2012da,Deldebbio:2010ze}. Without chiral symmetry breaking, this is also the more adequate form  \cite{Miransky:1986ib,Gusynin:1986fu}.

Of course, given that $b\approx 0$ in \pref{qfit} for small quark masses, it is possible to identify the exponents of the logarithms, $\gamma_p=1-\gamma$. This would yield then a $\gamma_p$ close to zero, and therefore smaller than other determinations. Given this relevance of the fit form, this should be regarded as a systematic error. Hence, the result is, given all the assumptions, that $\gamma\approx 1$ if chiral symmetry is spontaneously broken in the chiral limit, and $\gamma\le 0.3$ if not.

\subsection{Results in position space}\label{sups}

\begin{figure}
\includegraphics[width=\linewidth]{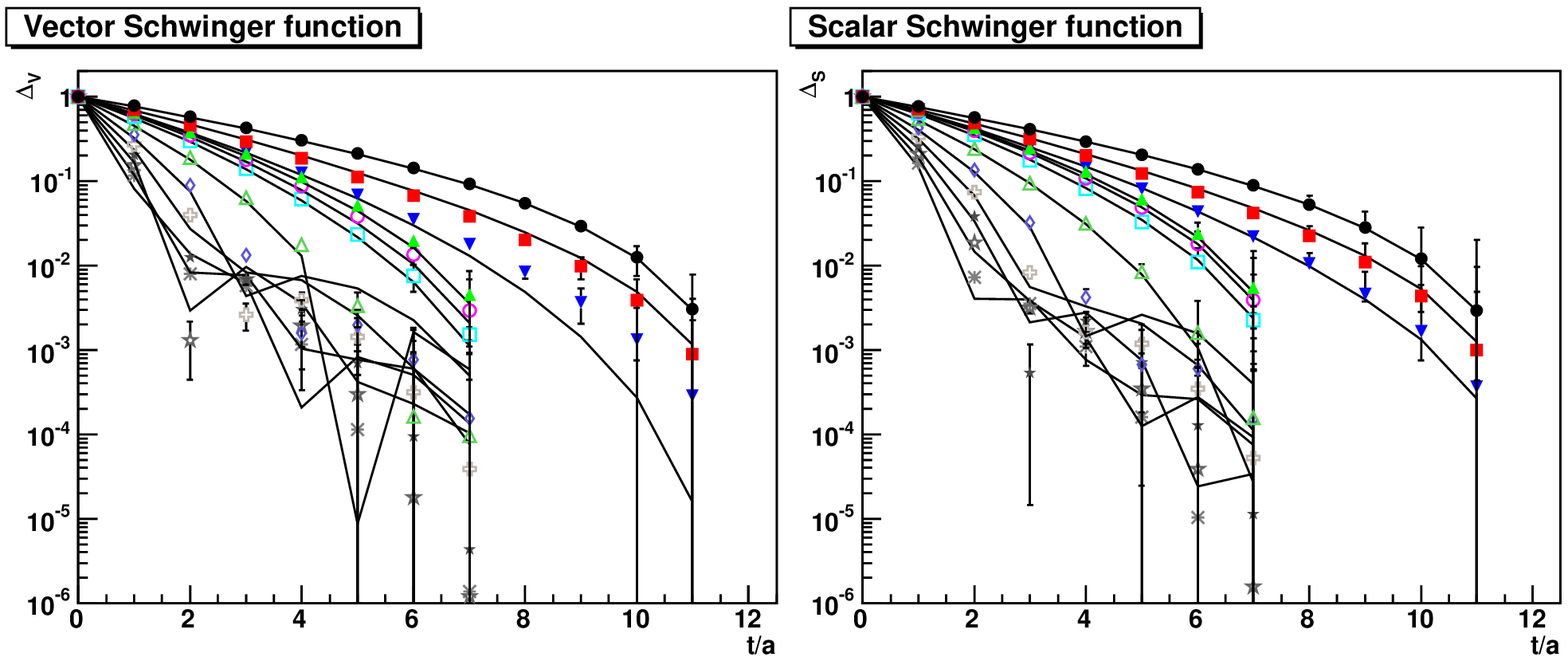}
\includegraphics[width=\linewidth]{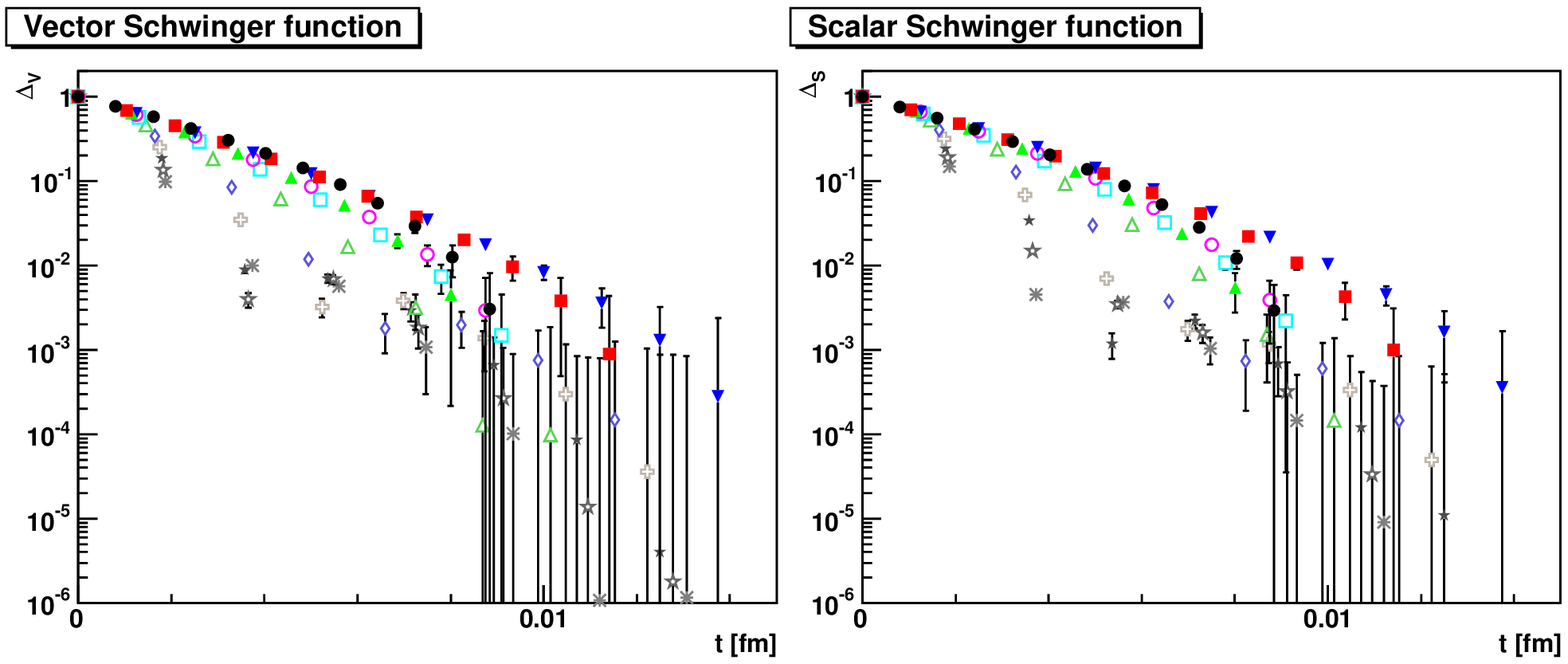}
\caption{\label{fig:tcp}The vector (left panels) and scalar (right panels) Schwinger functions in lattice units (top panels) and physical units (bottom panels) for different masses and volumes. For the case of lattice units also fits of type \pref{qfit} are shown. Renormalization is performed at $a\mu=1$. The symbols are the same as in figure \ref{fig:tcm}.}
\end{figure}

The Schwinger functions, shown in figure \ref{fig:tcp}, show no significant changes, compared to the quenched case. The bending in all cases shows that the dynamical adjoint quarks do not posses a positive-definite spectral function, and therefore do not belong to the physical state space. At larger masses, the zero crossing is still explicitly observable, but this is no longer possible at small quark masses. Also, in both lattice and physical units, the position of the first zero crossing moves towards larger times the smaller the mass.

Though the violation of positivity is not sufficient for confinement \cite{Maas:2011se}, it implies that the quarks are not observable particles. Thus,they must be either bound in bound states or confined, for all the masses investigated here. Furthermore, the behavior is neither a (approximate) power-law nor close to one, as would be expected if the theory would be (quasi-)conformal \cite{DiFrancesco:1997nk}. In fact, the fits of type \pref{qfit} describe the Schwinger functions quite well.

Combining both results, speculatively the behavior of the adjoint quarks in the limit of zero quark mass is best characterized as that of confined quarks without spontaneous chiral symmetry breaking. This underlines that adjoint quarks are substantially different from fundamental quarks.

\section{Summary}\label{ssummary}

Summarizing, the investigation of the adjoint quark propagator presented provided a number of interesting observations, though further systematic studies are necessary for any final statements. In any case, adjoint quarks are quite different from fundamental quarks. Especially their wave-function renormalization appears to be essentially identical to one, in strong contrast to the fundamental case \cite{Fischer:2006ub,Kamleh:2007ud}. Furthermore, in the chiral limit the results suggest the possibility that spontaneous chiral symmetry breaking vanishes in the chiral limit, at least for 2 colors and 2 flavors. At the same time the Schwinger functions show clearly a violation of positivity. Thus adjoint quarks do not belong to the physical spectrum, irrespective of whether in the quenched case or not. Finally, the mass function can be rather well described by the form \pref{qfit}. The interpretation in terms of an anomalous dimensions hinges crucially on the presence of spontan!
 eous chiral symmetry breaking in the chiral limit. If this breaking is absent, the anomalous dimension is small, in agreement with other determinations, otherwise it is large, close to one. Speculatively, this and other indications presented here, could be taken as a sign for the absence of chiral symmetry breaking in the chiral limit, in line with \cite{Sannino:2009za,DelDebbio:2010hu,DelDebbio:2010hx,Bursa:2009we,DelDebbio:2009fd,Catterall:2008qk,Catterall:2007yx,Hietanen:2008mr,Hietanen:2009az,DeGrand:2009mt,DeGrand:2011qd,Lucini:2009an,Catterall:2009sb,Maas:2011jf,Patella:2012da,Andersen:2011yj,Hill:2002ap,Lane:2002wv}. However, a much larger range of momenta would be necessary to be sure to have eliminated all residual systematic uncertainties introduced in the fitting process. Thus again a significantly enlarged investigation is called for.

In total, the results clearly show that adjoint quarks and fundamental quarks are very different, already for the quenched case, and become even more different towards the chiral limit. But much more systematic investigations will be needed to provide the necessary systematic reliability to make this a final statement. To really understand the chiral limit will probably require a chiral extrapolation of the dynamical case, which can be achieved using, e.\ g., functional methods, like done already for fundamental quarks \cite{Fischer:2006ub}.\\

\no{\bf Acknowledgments}

We are grateful to the authors of the papers \cite{DelDebbio:2010hu,DelDebbio:2010hx,DelDebbio:2009fd,DelDebbio:2008zf} for providing us with the unquenched configurations, making this investigation possible at all. This work was supported by the DFG under grant number MA 3935/5-1. Simulations were performed on the HPC cluster at the University of Jena. We are grateful to the HPC team for the very good performance of the cluster. The ROOT framework \cite{Brun:1997pa} has been used in this project.

\appendix

\bibliographystyle{bibstyle}
\bibliography{bib}


\end{document}